\documentclass[aps,twocolumn,showpacs,superscriptaddress,amsmath,amssymb,nofootinbib]{revtex4-1}
\usepackage{epsf,amsmath,amssymb,amsfonts,verbatim,color,multirow,pifont}
\usepackage{graphicx}
\usepackage{dcolumn}% Align table columns on decimal point
\usepackage{bm}% textbf math
\usepackage{txfonts}
\usepackage{hyperref}
\usepackage{soul}

\begin{document}
\title{Nonuniversality of heat engine efficiency at maximum power}

\author{Sang Hoon Lee}
%\email{lshlj82@kias.re.kr}
\affiliation{School of Physics, Korea Institute for Advanced Study, Seoul 02455, Korea}
\affiliation{Department of Liberal Arts, Gyeongnam National University of Science and Technology, Jinju 52725, Korea}

\author{Jaegon Um}
\email{slung@postech.ac.kr}
\affiliation{Quantum Universe Center, Korea Institute for Advanced Study, Seoul 02455, Korea}
\affiliation{CCSS, CTP and Department of Physics and Astronomy, Seoul National University, Seoul 08826, Korea}
\affiliation{BK21PLUS Physics Division, Pohang University of Science and Technology, Pohang 37673, Korea}

\author{Hyunggyu Park}
\email{hgpark@kias.re.kr}
\affiliation{School of Physics, Korea Institute for Advanced Study, Seoul 02455, Korea}
\affiliation{Quantum Universe Center, Korea Institute for Advanced Study, Seoul 02455, Korea}

%%%%%%%%%%

\begin{abstract}
We study the efficiency of a simple quantum dot heat engine at maximum power.
In contrast to the quasi-statically operated Carnot engine whose efficiency reaches the theoretical maximum,
recent research on more realistic engines operated in a finite time has revealed other classes of efficiencies
such as the Curzon-Ahlborn efficiency maximizing the power. Such a power-maximizing efficiency has been argued
to be always the half of the maximum efficiency up to the linear order near equilibrium under the tight-coupling condition between thermodynamic fluxes.
We show, however, that this universality may break down for the quantum dot heat engine,
depending on the constraint imposed on the engine control parameters (local optimization), even though the tight-coupling condition remains satisfied.
It is shown that this deviation is critically related to the applicability of the linear irreversible thermodynamics.
%{\color{red}
%In particular, it happens when the thermodynamics fluxes
%vanish simultaneously not only in reversible limit, but also in
%irreversible limit, so that it cannot be described by simple linear irreversible thermodynamics formalism.
%As a result, we dismiss the notion of universal linear coefficient of the efficiency at the maximum power,
%and discuss the implication of such a result in terms of entropy production and irreversible thermodynamics.}

\end{abstract}

\pacs{05.70.Ln, 05.70.-y, 05.40.−a}
% PACS, the Physics and Astronomy % Classification Scheme.%%
%%05.70.Ln  Nonequilibrium and irreversible thermodynamics
%%05.70.-a	Thermodynamics
%%05.40.-a	Fluctuation phenomena, random processes, noise, and Brownian motion
%%05.20.-y	Classical statistical mechanics
%%89.70.-a	Information and communication theory

%\keywords{}
%Use showkeys class option if keyword %display desired

\maketitle

%%%%%%%%%%%%%

\section{Introduction}
\label{sec:introduction}

The efficiency of heat engines is a quintessential topic of thermodynamics~\cite{HuangBook}. In particular, an elegant formula expressed only by hot and cold reservoir temperatures for the ideal quasi-static and reversible engine coined by Sadi Carnot has been an everlasting textbook example~\cite{Carnot1824}. That ideal engine, however, is not the most efficient engine any more when we consider its power output (the extracted work per unit time), which has added different types of optimal engine efficiencies such as the Curzon-Ahlborn (CA) efficiency, $\eta_\text{CA}$, for some cases~\cite{Chambadal1957,Novikov1958,Curzon1975}. Following such steps, researchers have taken simple systems to investigate various theoretical aspects of underlying principles of macroscopic thermodynamic engine efficiency in details~\cite{VanDenBroeck2005,Esposito2009PRL,Hoppenau2013,Proesmans2015,Um2015,Holubec2015,JMPark2016,Ryabov2016,Shiraishi2016}.

Most of model studies have focussed on the efficiency at maximum power output, $\eta_\text{op}$, with respect to all control parameters (global optimization)~\cite{ChenYan1989,SchmiedlSeifert2008,Tu2008,Esposito2009EPL,Esposito2012}.
In these studies, an interesting universality is found for $\eta_\text{op}$ such that a group of models show the identical $\eta_\text{op}$, e.g.~$\eta_\text{op}=\eta_\text{CA}=1-\sqrt{1-\eta_{C}}$ ($\eta_C$: Carnot efficiency) and few others. The universality becomes more broad near equilibrium (small $\eta_C$). Understanding this universality was first embarked by Van den Broeck~\cite{VanDenBroeck2005},
with a general description in the context of linear irreversible thermodynamics. Using a linear perturbation theory from equilibrium with a temperature-gradient field and a single external force field, he showed that
$\eta_\text{op}\approx \frac{1}{2} \eta_C$ up to the linear order of $\eta_C$ in case of the so-called tight coupling. Later on, this universality is extended to include the second order term such as $\frac{1}{8}\eta_C^2$~\cite{SchmiedlSeifert2008,Tu2008}, but this one is not as solid as the linear term~\cite{Esposito2009PRL,Esposito2010PRL,SSheng2014,SSheng2015}.

Van den Broeck's results do not demand global optimization of all control parameters, but only needs a single parameter varied with other parameters fixed (local optimization). Thus, the universality in the linear order
(called as the $\frac{1}{2}$-universality) does not discern local or global optimizations. In fact,
one can easily infer this from previous model studies~\cite{SSheng2014,SSheng2015}, where the $\frac{1}{2}$-universality is robust with non-universal second-order coefficients in local optimizations.

\begin{figure}
\includegraphics[width=\columnwidth]{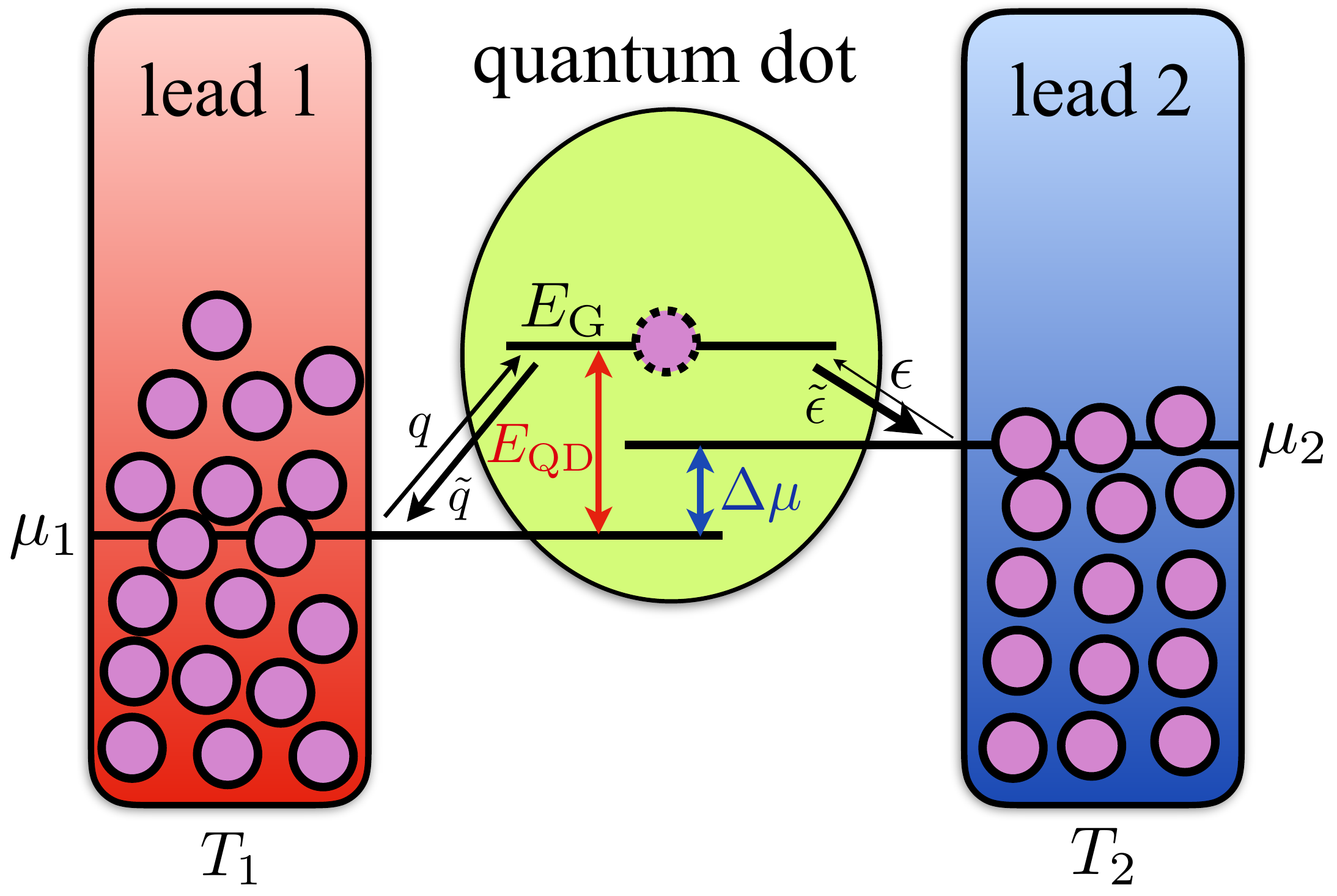}
\caption{A schematic illustration of the quantum dot heat engine. The quantum dot with a single energy level $E_\mathrm{G}$ is in contact with the leads, which plays the role of heat and particle reservoirs with temperatures $T_1$ and $T_2$, and chemical potentials $\mu_1$ and $\mu_2$. The maximum (Carnot) efficiency is given as $\eta_{C}=1-T_2/T_1$.
}
\label{fig:QD_engine_schematic}
\end{figure}

In this paper, we revisit the well-studied  quantum dot heat engine model~\cite{Esposito2009EPL,Esposito2012,Toral2016,exp} and consider various kinds of local optimizations.
The quantum dot engine is composed of a single quantum dot connected to two leads with characteristic temperatures and chemical potentials (Fig.~\ref{fig:QD_engine_schematic}).
We find an intriguing result: When the quantum dot energy level relative to one of the lead's chemical potential is fixed and the other is varied, the $\frac{1}{2}$-universality is observed with non-universal second-order coefficient depending on the value of the fixed chemical potential, as expected.
 On the other hand, when the quantum dot energy level is varied with fixed chemical potentials of both leads, we find that the robust $\frac{1}{2}$-universality is violated such that the linear coefficient turns out to be unity ($\eta_\text{op}\approx \eta_{C}$). This implies a much higher
efficiency at maximum power, compared to the conventional cases.

We emphasize that our engine always satisfies the tight-coupling condition in the sense that the heat flux is directly
proportional to the work-generating flux~\cite{VanDenBroeck2005,Esposito2009PRL}. This means that the universality
requires an additional constraint besides the tight-coupling condition, which turns out to be the applicability of the linear
irreversible thermodynamics~\cite{VanDenBroeck2005,Groot,SSheng2014,SSheng2015,Apertet,Gaveau,Izumida}. We point out that the latter non-universal case is also experimentally realizable as it corresponds to tuning the gate voltage of the quantum dot to optimize the power~\cite{exp}, while the control of the chemical potential difference of the leads can be done by adjusting the source-drain voltage~\cite{Kouwenhoven1997,YSLiu2013,Humphrey2002,Jordan2013}. In a recent experiment controlling the gate voltage, much higher efficiency than the usual CA efficiency was reported at maximum power~\cite{exp}, which supports our result.

The rest of the paper is organized as follows. We introduce the autonomous quantum dot heat engine model and its mathematically equivalent non-autonomous two-level model in Sec.~\ref{sec:model}. First, the global optimization of power in the entire parameter space is presented in Sec.~\ref{sec:series_expansion}. In Secs.~\ref{sec:optimizing_wrt_single_parameter} and \ref{sec:optimizing_wrt_alpha}, we present our main results for the optimization with various constraints and discuss its non-universal feature in power-maximizing efficiency. We conclude with the summary and a remark on future work in Sec.~\ref{sec:conclusion}.

\section{Heat engine models}
\label{sec:model}

\subsection{Quantum dot heat engine model}
\label{sec:QD_model}

We take a quantum dot heat engine introduced in Ref.~\cite{Esposito2009EPL}, which is composed of a quantum dot whose energy level $E_\mathrm{G}$ is controlled by the gate voltage where a single electron can occupy, in contact with two leads, denoted by $R_1$ and $R_2$ at different temperatures ($T_1 > T_2$) and chemical potentials ($\mu_1 < \mu_2 < E_\mathrm{G}$), respectively, as shown in Fig.~\ref{fig:QD_engine_schematic}.
For notational convenience, we define the energy level of the quantum dot as $E_\mathrm{QD} \equiv E_\mathrm{G} - \mu_1 >0$
and the chemical potential difference as $\Delta \mu = \mu_2 - \mu_1>0$.
Experimentally, it is possible to control $E_\mathrm{QD}$ by tuning the gate voltage connected to the quantum dot and
$\Delta\mu$ by tuning the source-drain voltage connected to the leads~\cite{Kouwenhoven1997}.

The transition rates of the electron to the quantum dot from $R_1$ and $R_2$ are given as the following Arrhenius form,
\begin{equation}
\begin{aligned}
q/\tilde{q} = e^{-E_\mathrm{QD}/T_1} \,, \\
\epsilon/\tilde{\epsilon} = e^{-(E_\mathrm{QD}-\Delta \mu)/T_2} \,,
\end{aligned}
\label{eq:q_and_epsilon}
\end{equation}
with $q$ ($\epsilon$) from $R_1$ ($R_2$) to the quantum dot and $\tilde{q}$ ($\tilde\epsilon$) vice versa.
Here, we set the Boltzmann constant $k_B = 1$.
Denoting the probability of occupation in the quantum dot by $P_{o}$ and its complementary probability (of absence) by $P_{e} = 1 - P_{o}$,
the probability vector $|\mathbf{P} \rangle = \left( P_{o} , P_{e} \right)^T$ is described by the master equation
\begin{equation}
\frac{d|\mathbf{P}\rangle}{dt} =
\begin{pmatrix}
-\tilde{q}-\tilde{\epsilon} & q+\epsilon \\
\tilde{q}+\tilde{\epsilon} & -q-\epsilon \end{pmatrix}
|\mathbf{P}\rangle \,.
\label{eq:probability_master_equation}
\end{equation}

For simplicity, tunneling rates between the quantum dot and the leads are chosen as $q + \tilde{q} = \epsilon + \tilde{\epsilon} = 1$.
Generalization to arbitrary finite rates does not change our main conclusions. With these normalized rates,
we find the condition for parameters as $0 \le \epsilon,\ q \le 1/2$.
The steady-state solution is easily obtained as
\begin{equation}
P_{o,ss}  = \frac{1}{2} \left( q + \epsilon \right) \,,\quad
P_{e,ss}  = \frac{1}{2} \left( 2 - q - \epsilon \right) \,,
\label{eq:master_equation_solution}
\end{equation}
with
\begin{equation}
E_\mathrm{QD}  = T_1 \ln\left[\left(1-q\right)/q\right] \,,\
E_\mathrm{QD}-\Delta\mu  = T_2 \ln\left[\left(1-\epsilon\right)/\epsilon\right] \,.
\label{eq:Eqeps}
\end{equation}
The probability currents from $R_1$ to the quantum dot and that from the quantum dot to $R_2$ are then,
\begin{equation}
\begin{aligned}
I_1 & = P_{e,ss} q - P_{o,ss} (1-q) = \frac{1}{2} (q-\epsilon) \,,\\
I_2 & = P_{o,ss} (1-\epsilon) - P_{e,ss} \epsilon = \frac{1}{2} (q-\epsilon) \,,
\end{aligned}
\label{eq:probability_current}
\end{equation}
respectively, and they are identical to each other, which represents the conservation of the particle flux. From now on, we denote
this steady-state particle flux carrying the energy current by
\begin{equation}
J \equiv \frac{1}{2} (q-\epsilon) \,.
\label{eq:J_definition}
\end{equation}

The heat production rate to the quantum dot from $R_1$ and that from the quantum dot to $R_2$ are
\begin{equation}
\begin{aligned}
\dot{Q}_1 & = J E_\mathrm{QD} \,, \\
\dot{Q}_2 & = J \left(E_\mathrm{QD}-\Delta\mu\right) \,.
\end{aligned}
\label{eq:Q1_and_Q2}
\end{equation}
A particle moving from the hot lead $R_1$ to the cold lead $R_2$ gains the energy $\Delta\mu$, which can be used later
as work against an external device. Thus, the {\em idealized} power of the engine is defined as
\begin{equation}
\dot{W} = \dot{Q}_1 - \dot{Q}_2 = J \Delta\mu \,,
\label{eq:Wnet}
\end{equation}
by the first law of thermodynamics. With $\Delta \mu >0$, we need the condition of $q \ge \epsilon$ (non-negative $J$) for a proper heat engine. It is clear that the tight coupling condition is satisfied in our model because the heat currents are proportional
to the work current with proportionality constants given by the non-vanishing ratio of energy control parameters.

The efficiency of the engine is given by the ratio
\begin{equation}
\eta = \frac{\dot{W}}{\dot{Q}_1} = \frac{\Delta\mu}{E_\mathrm{QD}} = 1 - \frac{T_2 \ln\left[\left(1 - \epsilon\right)/\epsilon\right]}{T_1 \ln\left[\left(1 - q\right)/q\right]} \,,
\label{eq:eta}
\end{equation}
which is independent of temperatures and the particle flux $J$.
By adjusting temperatures to approach the limit of $\epsilon\rightarrow q$ from below,
$\eta$ can reach the maximum (Carnot) efficiency~\cite{HuangBook,Carnot1824},
\begin{equation}
\eta_C = 1 - \frac{T_2}{T_1} \,.
\label{eq:Carnot_efficiency}
\end{equation}
The total entropy production rate in the steady state is given by the net entropy change rate of the leads;
\begin{equation}
\dot{S} = - \frac{\dot{Q}_1}{T_1} + \frac{\dot{Q}_2}{T_2} \ge 0\,.
\label{eq:entropy_change_one_cycle}
\end{equation}

\subsection{Cyclic two-level heat engine model}
\label{sec:TL_model}

\begin{figure}
\includegraphics[width=0.7\columnwidth]{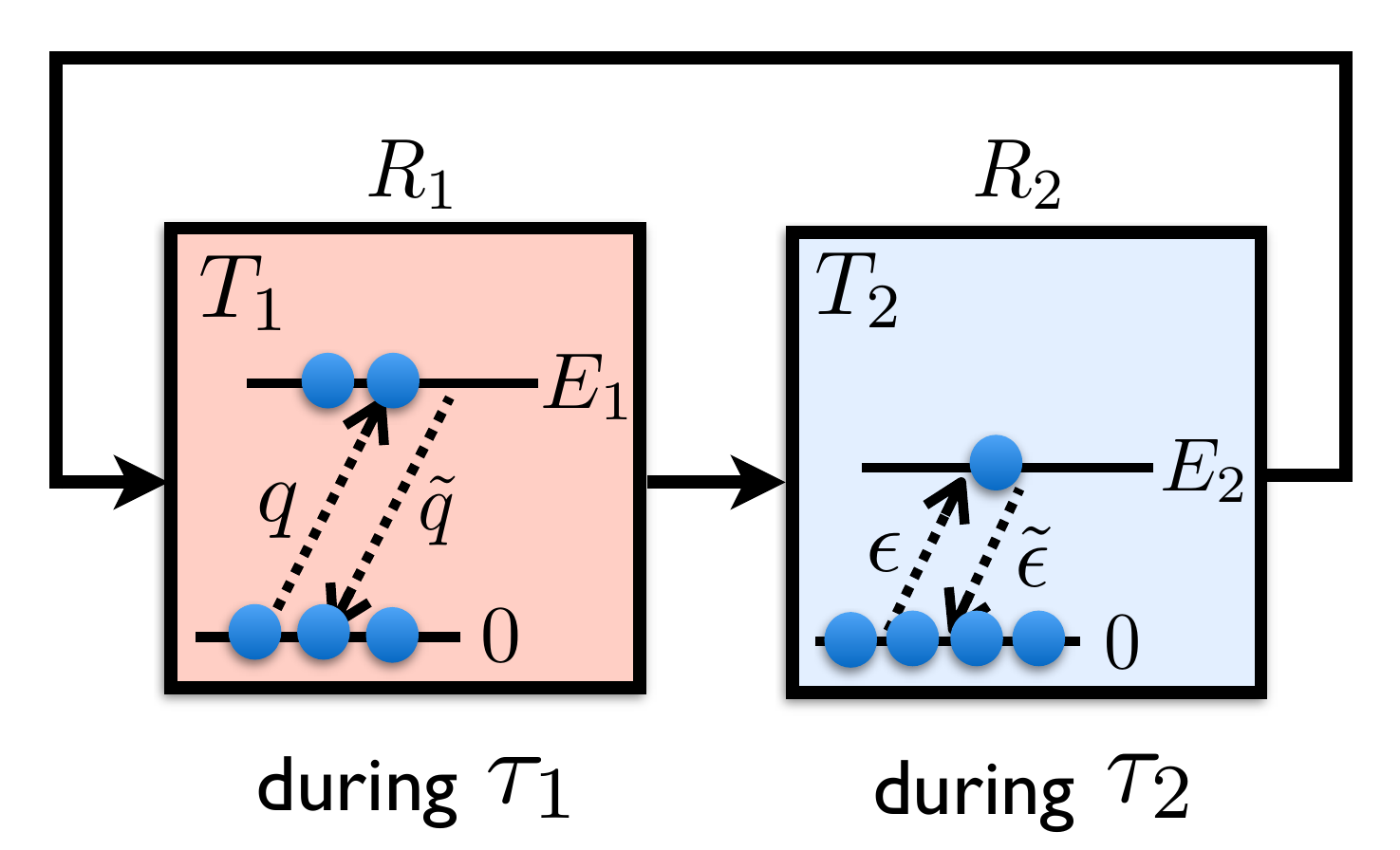}
\caption{Schematic illustration of a simple two-level heat engine, composed of two energy levels coupled with two heat reservoirs $R_1$ and $R_2$.}
\label{fig:schematic}
\end{figure}

The autonomous quantum dot heat engine introduced in Sec.~\ref{sec:QD_model} is in fact equivalent to a simple non-autonomous
cyclic two-level heat engine described in Fig.~\ref{fig:schematic}.
The two-level system is characterized by two discrete energy states composed of the ground state ($E = 0$) and the excited state ($E = E_1$ or $E = E_2$, depending on the contacting reservoir). The transition rates from the ground state to the excited state are denoted by $q$ and $\epsilon$, respectively, and their reverse processes by $\tilde{q}$ and $\tilde{\epsilon}$. We assume $E_1 > E_2$ and $T_1 > T_2$.

The system is attached to two different reservoirs: $R_1$ with temperature $T_1$ during time $\tau_1$, and $R_2$ with temperature $T_2$ during time $\tau_2$, and the adiabatic work extraction and insertion occur in between. Although the amount of energy unit involving the work exchange is the same ($E_1 - E_2$), the net positive work is achievable due to the difference in the population of the excited states at the end of contact with $R_1$ and $R_2$, which is determined by model parameters. Then, the mathematical formulation is exactly the same as the quantum dot engine if we use the following mapping from the energy variables in the quantum dot engine in Sec.~\ref{sec:QD_model}:
\begin{equation}
E_1 \equiv E_\mathrm{QD} \,,
\quad
E_2 \equiv E_\mathrm{QD} - \Delta \mu \,.
\label{eq:E_2_definition}
\end{equation}

Using the same formalism as in the autonomous quantum dot engine except for the explicit time dependence, the net work per cycle
(cyclic period $\tau=\tau_1+\tau_2$) in the cyclic steady state is given as
\begin{equation}
W_\textrm{net,two-level}  = \frac{\left(1-e^{-\tau/2}\right)^2 \left(q-\epsilon\right)}{1-e^{-\tau}}
\left(E_1 - E_2\right) \,,
\label{eq:Wnet_for_tau}
\end{equation}
assuming $\tau_1 = \tau_2 = \tau / 2$ for simplicity. The $\tau$-dependent factor $\left(1-e^{-\tau/2}\right)^2/\left(1-e^{-\tau}\right)$
is decoupled from the rest of the formula and thus is just an overall factor. The decoupling holds regardless of the  $\tau_1 = \tau_2$ condition; the overall factor becomes $\left(1-e^{-\tau_1}\right)\left(1-e^{-\tau_2}\right)/\left[1-e^{-(\tau_1+\tau_2)}\right]$. The mean power is then given by
\begin{equation}
\frac{W_{\textrm{two-level}}}{\tau} = \frac{\left(1-e^{-\tau/2}\right)^2 \left(q-\epsilon\right)}{\tau \left(1-e^{-\tau}\right)}
(E_1 - E_2) \,,
\label{eq:power_for_two_level}
\end{equation}
which decreases monotonically with $\tau$.

This result is exactly the same as the power for the quantum dot engine in
Eq.~\eqref{eq:Wnet} by replacing the current $J$ by $J_\text{cyc}=a(\tau) J$ with $a(\tau)=2(1-e^{-\tau/2})^2 / [\tau(1-e^{-\tau})]$.
In fact, all formulas for various other quantities are also written with $J_\text{cyc}$ instead of $J$, thus the analysis for the quantum dot engine
in the following sections should apply to the cyclic two-level heat engine with a trivial overall factor $a(\tau)$.
We remark that this kind of equivalence between the autonomous and the cyclic model does not exist for multi-level quantum dot models~\cite{LUP2016}.

\section{Efficiency at maximum power: global optimization}
\label{sec:series_expansion}

In this section, we investigate the efficiency at maximum power for the quantum dot engine.
We rewrite Eq.~(\ref{eq:Wnet}) for power in terms of $q$ and $\epsilon$ as
\begin{equation}
\dot{W}(q, \epsilon) = \frac{1}{2}\left(q-\epsilon \right)
\left[T_1 \ln \left( \frac{1-q}{q} \right) - T_2 \ln \left( \frac{1-\epsilon}{\epsilon} \right)  \right].
\label{eq:power}
\end{equation}
The condition for a proper heat engine with non-negative power ($\dot{W} \ge 0$) further restricts the parameter space of $(q,\epsilon)$
with
\begin{equation}
 \frac{1-q}{q} \le \frac{1-\epsilon}{\epsilon} \le \left(\frac{1-q}{q}\right)^{T_1/T_2}   \,.
\label{eq:restricted}
\end{equation}
Note that the lower bound corresponds to $J=0$ (the {\em reversible} limit with $\dot{S}=0$)
and the upper bound corresponds to $\Delta \mu =0$ (no work limit with $\dot{W}=0$).

For given $T_1$ and $T_2$, the power can be maximized at $(q^*,\epsilon^*)$ inside the above restricted parameter space,
which satisfies
\begin{equation}
\displaystyle \left. \frac{\partial \dot{W}}{\partial q} \right|_{(q^*,\epsilon^*)}
= \left. \frac{\partial \dot{W} }{\partial \epsilon} \right|_{(q^*,\epsilon^*)} = 0  \,,
\label{eq:first_derivative_condition}
\end{equation}
(see details in Appendix~\ref{sec:series_expansion_procedure}).
The efficiency at maximum power, $\eta_\mathrm{op}$, can be obtained from Eq.~\eqref{eq:eta} with $(q^*,\epsilon^*)$, which
is a function of the temperature ratio $T_2/T_1 (=1-\eta_C)$. This function cannot be written in a closed form with $\eta_C$, but
its expansion near equilibrium (small $\eta_C$) is given by
\begin{equation}
\eta_\mathrm{op} = \frac{1}{2} \eta_C + \frac{1}{8} \eta_C^2 + \frac{7-24a_0+24a_0^2}{96(1-2a_0)^2} \eta_C^3 +
\mathcal{O}\left({\eta_C^4}\right) \,,
\label{eq:eta_op_expansion_wrt_eta_C_as_CA_main_text}
\end{equation}
with $a_0=q^*|_{\eta_C=0^+}\approx 0.083\,222$, which is the solution of
$2/(1-2a_0) = \ln [ (1-a_0)/a_0]$.  The same expression was reported previously in equivalent models~\cite{Esposito2009EPL,Toral2016}.

\begin{figure}
\includegraphics[width=\columnwidth]{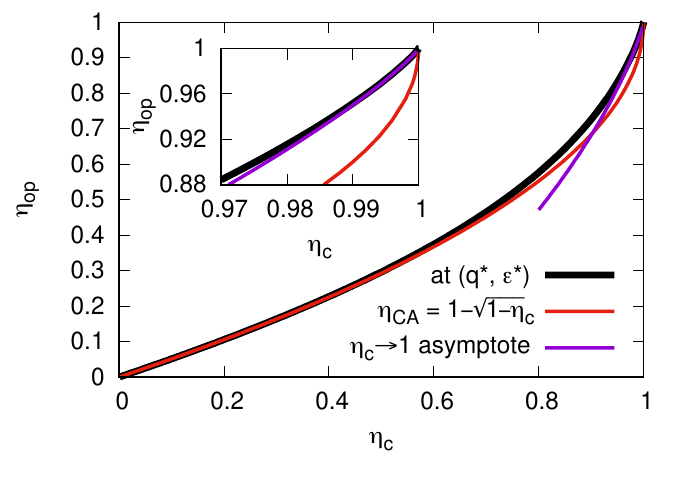}
\caption{The efficiency at maximum power $\eta_\mathrm{op}$ with respect to both $q$ and $\epsilon$ versus the Carnot efficiency $\eta_C$. The CA efficiency $\eta_\mathrm{CA}$ in Eq.~\eqref{eq:eta_CA_main_text} and
and the $\eta_C \to 1$ asymptote in Eq.~\eqref{eq:eta_op_expansion_near_eta_C_1_main_text} are also shown.
The inset shows the magnified view of the region $0.97 < \eta_C < 1$.
}
\label{fig:optimal_eta}
\end{figure}

We can compare $\eta_\mathrm{op}$ with the conventional Curzon-Ahlborn (CA) efficiency~\cite{Chambadal1957,Novikov1958,Curzon1975} as
\begin{equation}
\eta_\mathrm{CA} = 1 - \sqrt{T_2/T_1} = 1 - \sqrt{1 - \eta_C} \,,
\label{eq:eta_CA_main_text}
\end{equation}
with the expansion form
\begin{equation}
\eta_\mathrm{CA} = \frac{1}{2} \eta_C + \frac{1}{8} \eta_C^2 + \frac{1}{16} \eta_C^3 + \frac{5}{128} \eta_C^4 + \mathcal{O}(\eta_C^5) \,.
\label{eq:eta_CA_expansion_main_text}
\end{equation}
The two efficiencies share the same coefficients up to the quadratic terms in the expansion, which
are known to be {\em universal} due to tight-coupling between thermodynamic fluxes and the left-right
symmetry~\cite{VanDenBroeck2005,Esposito2009PRL}. The third order coefficient ($\simeq 0.077\,492$) in Eq.~\eqref{eq:eta_op_expansion_wrt_eta_C_as_CA_main_text}, however, is
different from $1/16$ ($=0.0625$) for the $\eta_\mathrm{CA}$.
Plots of $\eta_\text{op}$ and $\eta_\text{CA}$ against $\eta_C$
are shown in Fig.~\ref{fig:optimal_eta} for comparison.

The asymptotic behavior of $\eta_\mathrm{op}$ near $\eta_C = 1$ is given by
\begin{equation}
\eta_\mathrm{op} =  1 + (1-b_0) (1 - \eta_C) \ln (1 - \eta_C) + \mathcal{O}\left[ (1-\eta_C )\right] \,,
\label{eq:eta_op_expansion_near_eta_C_1_main_text}
\end{equation}
with $b_0=q^*|_{\eta_C=1^-}\approx 0.217\,812$ which is the solution of ${1}/(1-b_0)= \ln[(1-b_0)/b_0]$. This result is also shown in Fig.~\ref{fig:optimal_eta} for comparison with $\eta_\text{CA}$.

\section{Local optimization for one energy variable fixed}
\label{sec:optimizing_wrt_single_parameter}

For given $T_1$ and $T_2$, we fix the quantum dot energy  and one of the chemical potential.
We vary $\Delta \mu$ (thus $\epsilon$) with fixed $E_\mathrm{QD}$ (so fixed $q$).

\subsection{efficiency at maximum power}
\label{sec:power_maximization_fixed_q}

For given $q$ (or $E_\text{QD}$), we find $\epsilon^*$ maximizing the power in Eq.~\eqref{eq:power} with
$\partial \dot{W}/\partial \epsilon|_{\epsilon^*} = 0$ in the parameter space restricted
by Eq.~\eqref{eq:restricted}. A straightforward calculation similar to the global optimization in Sec.~\ref{sec:series_expansion}
yields the efficiency at maximum power for small $\eta_C$ as
\begin{equation}
\eta_\mathrm{op} =
\frac{1}{2} \eta_C + \frac{E_\mathrm{QD}}{16 T_2}
\tanh\left(\frac{E_\mathrm{QD}}{2T_2} \right) \eta_C^2  + \mathcal{O}\left({\eta_C^3}\right) \,.
\label{eq:optimal_eta_for_given_q}
\end{equation}
The linear coefficient $\frac{1}{2}$ may be regarded as natural due to the tight-coupling condition~\cite{VanDenBroeck2005} in our model.
More detailed discussion on this $\frac{1}{2}$-universality will be given later in Sec.\ref{sec:irreversible_thermodynamics_fixed_alpha}.

The quadratic coefficient is not universal, depending on the system parameter $E_\mathrm{QD}$, thus
differs in general from the universal value $\frac{1}{8}$ representing the left-right symmetry.
This implies that the left-right symmetry should be considered not only in the engine device by itself,
but also in the allowed parameter space which is broken in this local optimization case.
We note that the universal value
$\frac{1}{8}$ is recovered for the special case of
\begin{equation}
\frac{E_\mathrm{QD}}{ T_2} \tanh\left( \frac{E_\mathrm{QD}}{2T_2} \right) = 2 \, .
\label{eq:fixed_q_quadratic_condition}
\end{equation}

\begin{figure}
\includegraphics[width=0.9\columnwidth]{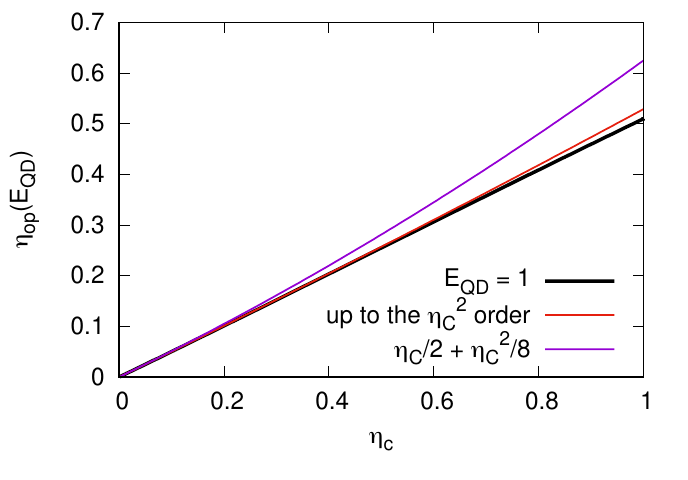}
\caption{The efficiency at maximum power $\eta_\mathrm{op}$ versus $\eta_C$ for $E_\mathrm{QD} = 1$ and $T_2 = 1$.
The black thick curve represents the exact one. The red thin line is drawn from the expansion
in Eq.~\eqref{eq:optimal_eta_for_given_q} up to the quadratic order, which is very close to the exact one.
 For comparison, we also plot the $\eta_C / 2 + \eta_C^2 / 8$ curve (purple thin line), which differs significantly.
}
\label{fig:optimal_eta_for_given_q}
\end{figure}

Plots of $\eta_\text{op}$ against $\eta_C$ are shown in Fig.~\ref{fig:optimal_eta_for_given_q}. It is interesting to note that
the asymptotic behavior of $\eta_\text{op}$ near $\eta_C=1$ is quite different from that in the case of the global optimization (see Sec.~\ref{sec:series_expansion}) and its leading order is given by
$\eta_\text{op}\approx 1-\alpha_0  + {\mathcal O}(1-\eta_C)$
with $\alpha_0$ satisfying the equation $1=\alpha_0 +(T_2/E_\text{QD})\sinh [\alpha_0E_\text{QD}/T_2]$. Note that
$0\le \alpha_0 \le 1/2$.

\subsection{irreversible thermodynamics approach}
\label{sec:irreversible_thermodynamics_fixed_q}

Near equilibrium, it is useful to analyze a heat engine in the viewpoint of irreversible thermodynamics~\cite{VanDenBroeck2005,SSheng2014,SSheng2015}.
The total entropy production rate in Eq.~\eqref{eq:entropy_change_one_cycle} can be written as
\begin{equation}
\dot{S}=\dot{Q}_1 \left(\frac{1}{T_2}-\frac{1}{T_1}\right) -\frac{\dot{W}}{T_2}
\equiv J_t X_t + J_1 X_1 \,,
\label{eq:entropy_for_irreversible_thermodynamics_q_fixed}
\end{equation}
with
the thermal flux
\begin{equation}
J_t = \dot{Q}_1 = J  E_\mathrm{QD}  \,,
\label{eq:thermal_flux_q_fixed}
\end{equation}
the thermal force representing the temperature gradient
\begin{equation}
X_t = \frac{1}{T_2} - \frac{1}{T_1} = \frac{\eta_C}{T_2} \,,
\label{eq:thermal_force_q_fixed}
\end{equation}
the mechanical flux
\begin{equation}
J_1 = - J T_2 \,,
\label{eq:mechanical_flux_q_fixed}
\end{equation}
and the mechanical force representing the chemical potential gradient,
\begin{equation}
X_1 = \frac{\Delta \mu}{T_2^2} \,.
\label{eq:mechanical_force_q_fixed}
\end{equation}
Accordingly, the product of mechanical flux and mechanical force leads to the power
\begin{equation}
\dot{W} = J \Delta \mu= - T_2 J_1 X_1 \,.
\label{eq:power_in_terms_of_J1_X1}
\end{equation}
The condition $X_t = X_1 = 0$ corresponds to the thermal and mechanical equilibrium state with $\dot{S}=\dot{W}=0$.

We expand the particle flux $J$ in Eq.~\eqref{eq:J_definition} for small forces $X_t$ and $X_1$ (small $\eta_C$ and $\Delta\mu$)
and find, after some algebra,
\begin{eqnarray}
J_1 &=& L \left( X_1 + \xi X_t \right) \left[  1+ \gamma \left( X_1 -\xi X_t \right) \right]
  + \mathcal{O}\left(X_t^3,X_1^3 \right) \label{eq:J_1}\\
J_t &=& \xi J_1\,,
\label{eq:J_t}
\end{eqnarray}
where
\begin{equation}
\label{eq:xi_form}
L = \frac{ T_2^2 e^{-E_\mathrm{QD}/T_2} } {2 \left( 1 + e^{-E_\mathrm{QD} /T_2} \right)^2}\,,
\xi =  - \frac{E_\mathrm{QD}}{T_2}\,, \gamma = \left( \frac{T_2}{2} \right) \tanh\left( \frac{E_\mathrm{QD}}{2T_2} \right).
\end{equation}
Eq.~\eqref{eq:J_t} indicates that the tight-coupling condition is satisfied~\cite{VanDenBroeck2005}.

We optimize power in Eq.~\eqref{eq:power_in_terms_of_J1_X1} with respect to $X_1$ and find
the optimal $X_1^*$ up to the quadratic order of $X_t$ as
\begin{equation}
X_1^* = - \frac{\xi}{2} X_t + \frac{\gamma \xi^2}{8} X_t^2\,.
\label{eq:X_m_star_as_X_t}
\end{equation}
Since the efficiency  is given by
\begin{equation}
\eta  =\frac{\dot{W}}{\dot{Q_1}}=- \frac{J_1 X_1 T_2}{J_t} = - \frac{X_1 T_2}{\xi} \,,
\label{eq:eta2}
\end{equation}
the efficiency at maximum power is obtained as
\begin{equation}
\eta_\mathrm{op} = \frac{1}{2} \eta_C -\frac{\xi \gamma}{8T_2} \eta_C^2 + \mathcal{O}\left( \eta_C^3 \right)  \,,
\label{eq:eta_op_for_X1}
\end{equation}
which is obviously the same as that in Eq.~\eqref{eq:optimal_eta_for_given_q}. The condition of Eq.~\eqref{eq:fixed_q_quadratic_condition}
to get the universal quadratic coefficient $\frac{1}{8}$ is equivalent to the ``energy-matching condition'' described in Ref.~\cite{SSheng2015}.

\section{Optimization for chemical potential difference fixed}
\label{sec:optimizing_wrt_alpha}

For given $T_1$ and $T_2$, we fix both chemical potentials and vary $E_\mathrm{QD}$ to find the power maximum.
This situation is natural and easily realizable experimentally for a quantum dot engine where the source-drain voltage is fixed, while the gate voltage is adjusted to maximize the power~\cite{Kouwenhoven1997,YSLiu2013,Humphrey2002,Jordan2013}. It is in contrast to the previous cases where the maximum power is obtained by adjusting either or both of the source-drain voltages.

\subsection{efficiency at maximum power}
\label{sec:power_maximization_fixed_alpha}

It is convenient to rewrite the expression for power in Eqs.~\eqref{eq:Wnet} and \eqref{eq:power} in terms of
energy variables $\Delta\mu$ and $E_\mathrm{QD}$ as
\begin{equation}
\dot{W} = \frac{1}{2}\left( \frac{e^{-E_\mathrm{QD}/T_1} }{1+e^{-E_\mathrm{QD}/T_1} }
- \frac{e^{-E_\mathrm{QD}/T_2} e^{\Delta\mu/T_2}}{1+e^{-E_\mathrm{QD}/T_2} e^{\Delta\mu/T_2}} \right) \Delta\mu \,.
\label{eq:power_in_terms_of_alpha}
\end{equation}
For fixed $\Delta\mu >0$, $\dot{W}$ varies with $E_\mathrm{QD}$ in the parameter range of  $E_\mathrm{QD} \ge \Delta\mu/\eta_C$ ($q\ge\epsilon$). Note that the boundary point $E_\mathrm{QD}^r = \Delta\mu/\eta_C$ is a reversible one, where $\dot{W}=\dot{S}=0$ along with
$\eta=\eta_C$.

As $E_\mathrm{QD}$ increases from the reversible value, $\dot{W}$ increases first but should decrease later after an optimal point because
Eq.~\eqref{eq:power_in_terms_of_alpha} indicates that $\dot{W}$ should vanish as $E_\mathrm{QD} \rightarrow\infty$.
The asymptotic point ($E_\mathrm{QD} =\infty$) is special with the particle current $J=0$ in Eq.\eqref{eq:J_definition} but
$q/\epsilon=e^{-\Delta\mu/T_2}\neq 1$ (broken detailed balance).
We call this asymptotic point as the {\em zero-flux nonequilibrium} point, in contrast to the ordinary
zero-flux equilibrium point ($J=0$ with detailed balance $q=\epsilon$).

The optimal point with maximum power is obtained by
\begin{equation}
\left. \frac{\partial \dot{W}}{\partial E_\mathrm{QD}}\right|_{E_\mathrm{QD} = E_\mathrm{QD}^*} = 0 \,,
\label{eq:power_optimize_wrt_EQ}
\end{equation}
where the optimal $E_\mathrm{QD}^*$ satisfies
\begin{equation}
\frac{e^{-E_\mathrm{QD}^*/T_1}}{\left( 1 + e^{-E_\mathrm{QD}^*/T_1} \right)^2} \frac{T_2}{T_1}
= \frac{e^{-E_\mathrm{QD}^*/T_2} e^{\Delta\mu/T_2}}{\left( 1 + e^{-E_\mathrm{QD}^*/T_2}e^{\Delta\mu/T_2} \right)^2} \,.
\label{eq:EQ_star_equation}
\end{equation}

\begin{figure}
\includegraphics[width=0.9\columnwidth]{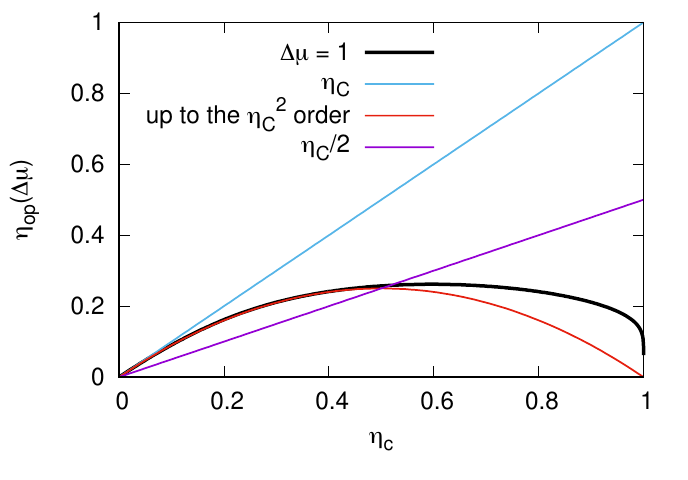}
\caption{The efficiency at maximum power $\eta_\mathrm{op}$ for $\Delta\mu = 1$ and $T_2 = 1$.
The black thick curve represents the exact one. It clearly shows that the slope for small $\eta_C$ is $1$, rather than $\frac{1}{2}$.
The red thin line is drawn from the expansion in Eq.~\eqref{eq:eta_op_for_given_deltamu} up to the quadratic order, which is very close
to the exact one up to $\eta_C\approx 0.4$.
}
\label{fig:eta_op_for_alpha_1p0_T2_1p0}
\end{figure}

First, consider the asymptotic behavior near small $\eta_C$. The reversible value $E_\text{QD}^r=\Delta\mu/\eta_C$ diverges as well as
the optimal point $E_\text{QD}^*$. Keeping the lowest order terms of $e^{-E_\text{QD}^*/T_2}$ in Eq.~\eqref{eq:EQ_star_equation},
we easily obtain
\begin{equation}
E_\mathrm{QD}^* = \frac{\Delta\mu}{\eta_C} - \frac{T_2}{\eta_C} \ln\left(1 - \eta_C\right).
\label{eq:EQ_star}
\end{equation}
Inserting this into  Eq.~\eqref{eq:eta},
we finally arrive at the efficiency at maximum power as
\begin{equation}
\eta_\mathrm{op}
= \eta_C - \frac{T_2}{\Delta\mu} \eta_C^2 + \mathcal{O}\left( \eta_C^3 \right) \,.
\label{eq:eta_op_for_given_deltamu}
\end{equation}
In contrast to the previous cases, the linear coefficient in the expansion deviates from the
$\frac{1}{2}$-universality and becomes unity along with the negative quadratic coefficient.  This example clearly illustrates that this seemingly robust $\frac{1}{2}$-universality for conventional tight-coupling engines~\cite{VanDenBroeck2005} can be also violated, depending on the type of restricted control-parameter spaces used in the power
maximization. In the next subsection, we will discuss about the violation of the $\frac{1}{2}$-universality in the perspective of irreversible thermodynamics
and the singular behaviors of thermodynamic and mechanical fluxes.

Next, we consider near $\eta_C\approx 1$. In Fig.~\ref{fig:eta_op_for_alpha_1p0_T2_1p0} where the exact result (numerically obtained) is displayed for all values of $\eta_C$, we note that $\eta_\text{op}$ does not increase monotonically with $\eta_C$ and
vanishes at $\eta_C=1$ with a singularity.  After some algebra, we find indeed a logarithmic singularity
such as $\eta_\text{op} \approx (\Delta\mu/T_2)/[-\ln(1-\eta_C)]$.

\subsection{irreversible thermodynamics approach}
\label{sec:irreversible_thermodynamics_fixed_alpha}

Now, we set up a perturbation theory near the zero-flux nonequilibrium point at $E_\text{QD}=\infty$
and $\eta_C=0$.
As $E_\text{QD}$ is varied with fixed $\Delta\mu$, the mechanical force
$X_1$ in Eq.~\eqref{eq:mechanical_force_q_fixed}
cannot be used as a mechanical force variable. Instead, we choose another mechanical force defined as
\begin{equation}
X_2 = \frac{1}{E_\mathrm{QD} } \,,
\label{eq:mechanical_force}
\end{equation}
which approaches the zero-flux nonequilibrium point as $X_2\rightarrow 0$
and has the dimension of inverse energy like $X_1$ and $X_t$.
The corresponding mechanical flux $J_2$ should be given as
\begin{equation}
J_2 = -\frac{\dot{W}}{T_2X_2}=
- J \frac{\Delta\mu}{T_2}  E_\mathrm{QD} \,,
\label{eq:mechanical_flux}
\end{equation}
which makes the entropy production rate in the standard form as
\begin{equation}
\dot{S} = J_t X_t + J_2 X_2 \,,
\label{eq:entropy_for_irreversible_thermodynamics_delta_mu_fixed}
\end{equation}
with the same thermal flux $J_t = J E_\text{QD}$ and thermal force $X_t = \eta_C/T_2$
in Eqs.~\eqref{eq:thermal_flux_q_fixed} and \eqref{eq:thermal_force_q_fixed}.

Notice that, at $X_2=0$,  the particle current $J$ vanishes (exponentially) as well as $\dot{Q}_1=\dot{W}=0$
with $\eta=\Delta\mu/E_\text{QD}=0$ (see Eq.~\eqref{eq:eta}), even with nonzero $X_t$.
This is the {\em crucial} difference with the previous case with fixed $E_\text{QD}$ in Sec.~\ref{sec:optimizing_wrt_single_parameter}, where both $X_1=0$ and $X_t=0$
are necessary to get all vanishing fluxes. The reason behind this difference is
{\em nonequilibrium-ness} of the $E_\text{QD}=\infty$ point, which does not necessarily obey
the standard linear response theory around equilibrium.

As mentioned in the previous subsection, the detailed balance is broken due to
$q/\epsilon \neq 1$. Even though $\dot{S}=0$ at this point, the average entropy production per
one particle transfer diverges as $\dot{S}/J \approx E_\text{QD}\eta_C/T_2$, which reveals its {\em irreversible} feature. (A similar situation was discussed in~\cite{JSLee2017}.)
Therefore, although our approach is dealing with vanishing fluxes in the limit of $X_t \to 0$ and $X_2 \to 0$, it is technically not the conventional irreversible thermodynamics used in Refs.~\cite{Groot,SSheng2014,SSheng2015},
which is a perturbation theory near  the true equilibrium state.
Nevertheless, in the following, we present the same type of irreversible thermodynamics analysis and
its implication for better understanding of the situation.

\begin{figure}
\begin{tabular}{l}
(a) \\
\includegraphics[width=0.9\columnwidth]{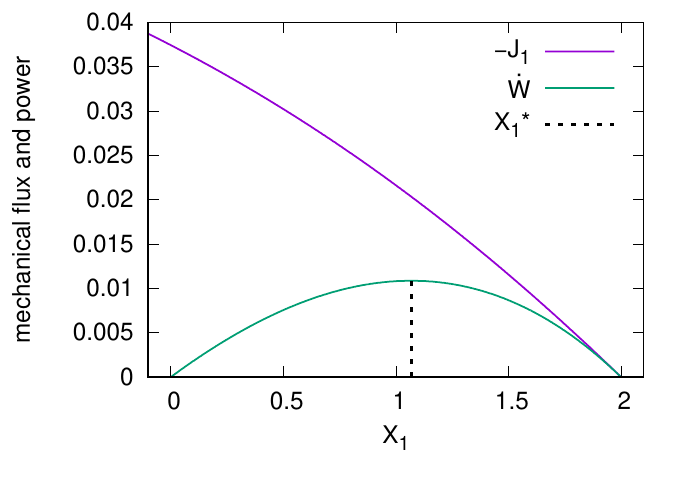} \\
(b) \\
\includegraphics[width=0.9\columnwidth]{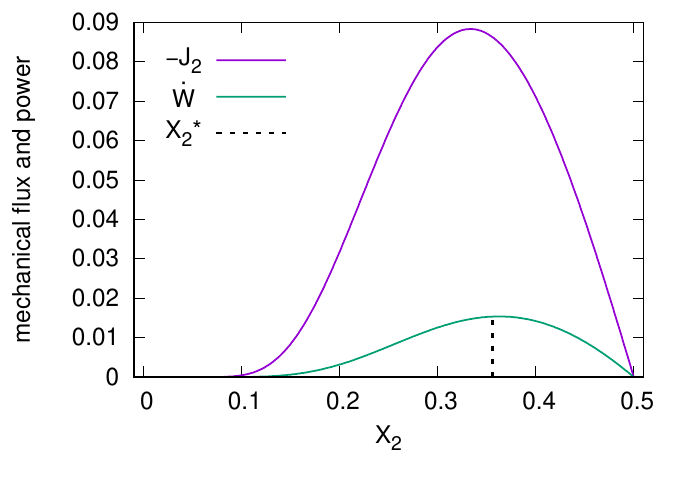} \\
\end{tabular}
\caption{Comparison between  (a) the fixed-$E_\mathrm{QD}$ case and  (b) the fixed-$\Delta\mu$ case, in terms of the mechanical flux (we plot the negative value of the flux for better visualization) and power,
where we set $T_1 = 1$ and $T_2 = 1/2$.
In (a), we plot $J_1$ and $\dot{W}$ against the mechanical force $X_1$ at $E_\mathrm{QD} = 1$. Both $J_1$ and $\dot{W}$ vanish at $X_1 = -\xi X_t = E_\mathrm{QD} \eta_C / T_2^2 = 2$, but only $\dot{W}$ vanishes at $X_1 = 0$. In (b), we plot $J_2$ and $\dot{W}$ against $X_2$ at $\Delta\mu = 1$. In this case, both $J_2$ and $\dot{W}$ vanish at both $X_2 = - \xi' X_t = \eta_C / \Delta\mu = 1/2$ and $X_2 = 0$.
For each case, we indicate the optimal values of $X_1^*$ and $X_2^*$ at maximum power.
}
\label{fig:J_vs_X}
\end{figure}

First, we again observe the tight-coupling condition between $J_t$ and $J_2$ from Eq.~\eqref{eq:mechanical_flux} as
\begin{equation}
J_t/J_2  = -T_2 /\Delta\mu \equiv \xi'  \,,
\label{eq:xip}
\end{equation}
which is a constant in the optimization process in this section.
This condition guarantees that the reversible condition $\dot{S}=0$ can be achieved at non-zero forces with $X_{2} = -\xi' X_t$,
similar to the standard irreversible thermodynamics discussed in Sec.~\ref{sec:irreversible_thermodynamics_fixed_q}.
Expansion of the mechanical flux in Eq.~\eqref{eq:mechanical_flux} for small forces $X_t$ and $X_2$ leads to
\begin{equation}
J_2 = \frac{\Delta \mu}{2T_2 X_2} e^{-\frac{1}{ T_2 X_2} }\left(
e^{\Delta \mu /T_2} -e^{X_t/X_2}  \right) + \mathcal{O}\left( e^{-\frac{2}{ T_2 X_2} },  e^{\frac{2(T_2 X_t-1)}{T_2 X_2}} \right)\,,
\label{eq:J_2_as_T_2_X_2}
\end{equation}
which vanishes as $X_2\sim X_t \rightarrow 0$ with an {\em essential} singularity rather than linearly seen in Sec.~\ref{sec:irreversible_thermodynamics_fixed_q}. This implies that the {\em linear} irreversible thermodynamic
analysis is not applicable to our case.

Plots of the power $\dot{W}$ and the mechanical flux $J_1$ for the fixed-$E_\text{QD}$ case in Sec.\ref{sec:power_maximization_fixed_q} and $J_2$ for the fixed-$\Delta\mu$ case in this section are shown for comparison in Fig.~\ref{fig:J_vs_X}.
For the fixed-$E_\text{QD}$ case shown in Fig.~\ref{fig:J_vs_X}(a), $\dot{W}$ should be approximated as a simple parabola for very small $\eta_C$
(thus very small parameter interval), because the limiting behaviors near both boundaries ($X_1=0$ and $X_1=-\xi X_t$) are linear,
which is usually the case in most optimization procedures.
Then the optimal $X_1^*$ is right at the middle point ($X_1^*=-\xi X_t/2$).
On the other hand, the efficiency increases linearly such as $\eta=\Delta\mu/E_\text{QD}\sim X_1$ and reaches $\eta_C$ at the
reversible point ($X_1=-\xi X_t$). Thus
we can easily expect the $\frac{1}{2}$-universality ($\eta_\text{op}\simeq\eta_C/2$) at maximum power, in general.

However, for the fixed-$\Delta\mu$ case, the functional behavior of $\dot{W}$ near $X_2=0$ is anomalous with an essential
singularity, seen in Eq.~\eqref{eq:J_2_as_T_2_X_2} and in Fig.~\ref{fig:J_vs_X}(b).  When the parameter interval becomes very small (small $\eta_C$),
one can easily expect the optimal $X_2^*$ should approach the reversible point $X_2=-\xi^\prime X_t$, leading to
$\eta_\text{op}\simeq \eta_C$ found in Eq.~\eqref{eq:eta_op_for_given_deltamu}.

For simple analysis, we consider a nonlinear leading term of an arbitrary order in the mechanical flux  as
\begin{equation}
J_2 = L' \left( X_2 + \xi' X_t \right) X_2^n \,,
\label{eq:J_m_as_X_m_and_X_t_extended_fixed_alpha}
\end{equation}
which vanishes at the reversible point $X_2=-\xi' X_t$.
For $n>0$, all fluxes ($J_2$ and $J_t$) vanish with $X_2=0$, regardless of the value of
$X_t$ (nonzero temperature gradient), which corresponds to our situation.

We optimize the power $\dot{W}=-T_2J_2 X_2$ in Eq.~\eqref{eq:mechanical_flux} with respect to
$X_2$ and find the optimal $X_2^*$ as
\begin{equation}
X_2^* = - \frac{n+1}{n+2} \xi' X_t \,,
\label{eq:X_m_star_as_X_t_fixed_alpha}
\end{equation}
and the efficiency at maximum power is obtained as
\begin{equation}
\eta_\mathrm{op} =  \frac{n+1}{n+2} \eta_C \,.
\label{eq:eta_op_for_a_fixed_alpha}
\end{equation}
The linear case ($n=0$) yields $\eta_{\rm op} = \eta_C/2$ for the tight-coupling heat engine~\cite{VanDenBroeck2005}
as expected.
However, our case with an essential singularity in Eq.~\eqref{eq:J_2_as_T_2_X_2} corresponds to the $n\rightarrow\infty$ limit,
leading to $\eta_\mathrm{op} \simeq \eta_C $, which is consistent with our result in Eq.~\eqref{eq:eta_op_for_given_deltamu}, up to the leading order. We remark that our heat engine provides only three possible values of the linear coefficient as $1/2$, $1$, and $0$
(varying $E_\text{QD}$ and $\Delta\mu$ together such as $\eta_C E_\text{QD}=\Delta\mu + b T_2$ with $b>0$).
We note that nonlinear response near equilibrium has been considered in a different context~\cite{WT,KI},
where a nonuniversal linear coefficient was also reported.

\subsection{Practical gain of the optimization with chemical potential difference fixed}
\label{sec:gain_fixed_alpha}

\begin{figure}
\begin{tabular}{l}
(a) \\
\includegraphics[width=0.9\columnwidth]{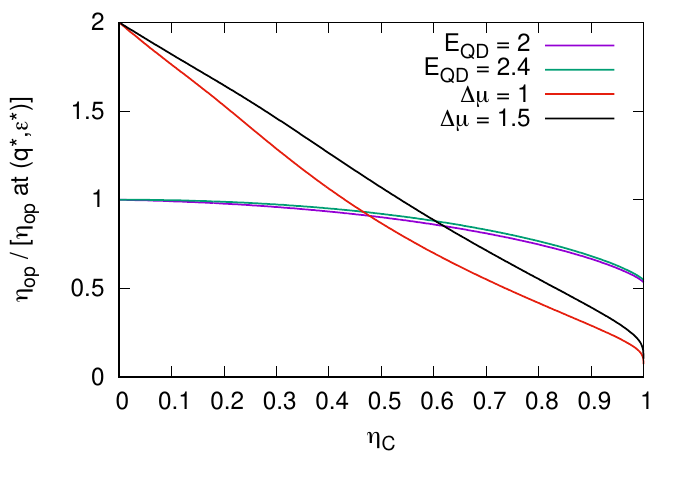} \\
(b) \\
\includegraphics[width=0.9\columnwidth]{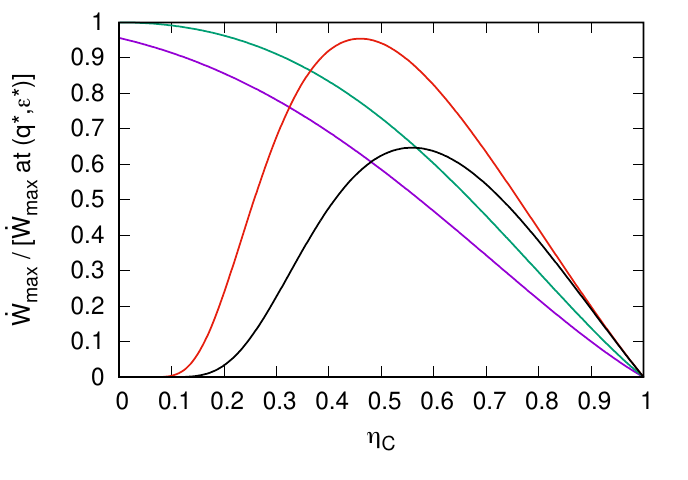} \\
\end{tabular}
\caption{Comparison between local and global optimizations.
(a) The efficiency at maximum power and (b) the maximum power in local optimizations scaled by those in global optimizations
with $T_2=1$.  The purple and green curves correspond to the fixed-$E_\mathrm{QD}$ case with
$E_\text{QD}=2$ and $2.4$. The red and black curves correspond to the fixed-$\Delta\mu$ case with
$\Delta\mu=1$ and $1.5$.
}
\label{fig:relative_efficiency_vs_max_power}
\end{figure}

The effectiveness of an engine should be featured by a high efficiency and a high power output.
However, there is a trade-off relation between the power and the efficiency~\cite{Shiraishi2016}, which
does not allow both merits simultaneously.  In previous subsections, we show that, for small $\eta_C$ (more realistic situations),
the power optimization with fixed $\Delta\mu$ provides us a higher efficiency at maximum power than that in the global optimization discussed in Sec.~\ref{sec:series_expansion}. But it is also obvious that its power output cannot be larger than that at the global maximum.

The efficiencies at maximum power $\eta_\text{op}$
in two local optimizations are shown in Fig.~\ref{fig:relative_efficiency_vs_max_power}(a) in comparison with that
in the global optimization. As expected, $\eta_\text{op}$ for the fixed-$\Delta\mu$ case is larger than that for the
global optimization for a rather wide range of $\eta_C$ ($\eta_C\lesssim 0.5$).
We also plot the maximum power in local optimizations scaled by the global optimum value
Fig.~\ref{fig:relative_efficiency_vs_max_power}(b). We note that the maximum power for the fixed-$\Delta\mu$ case
reaches up to a significant fraction of the global optimum value.
For example, the case of $\Delta\mu = 1$ at $\eta_C \simeq 0.3$ gives  about $30\%$ larger $\eta_\mathrm{op}$ than that for the global optimization case and reaches about $70\%$ of the global maximum power~\cite{exp1}. This engine at these parameter values may be viewed
as ``more effective'' than the globally optimized engine in some specific situations preferring a good efficiency.

\section{Conclusions and discussion}
\label{sec:conclusion}

We have demonstrated that a quantum dot heat engine exhibits various nonuniversal forms of the efficiency at maximum power $\eta_\text{op}$.
In particular, compared to the global or local optimization with varying source-drain voltages, the single-parameter optimization by controlling the gate voltage of the quantum dot for fixed source-drain voltages
reveals $\eta_\text{op}\approx \eta_C$ for small $\eta_C$, which breaks the $\frac{1}{2}$-universality
($\eta_\text{op}\approx \frac{1}{2}\eta_C$). This universality has been believed to be robust for any engine with the tight-coupling condition of thermodynamics fluxes.

We have investigated the origin of this universality break down in terms of irreversible thermodynamics and a singular behavior of the mechanical current. In fact, the absence of linear response regime of thermodynamic fluxes may yield various values of the linear coefficient in the standpoint of irreversible thermodynamics. In particular, the existence of a zero-flux nonequilibrium point is crucial.
Our case turns out to be an extreme case with an essential singularity in the mechanical current, which
makes the efficiency at maximum power close to the Carnot efficiency.
A recent experimental study for a quantum dot system~\cite{exp} shows results consistent with our theoretical finding. We expect the similar nonuniversality in the Smoluchowski Feynman ratchet model recently studied by one of us~\cite{JSLee2017}, where the equilibrium point is not accessible due to the existence of inherent irreversible heat currents.

The two mathematically identical two-level heat engine models (autonomous engine and non-autonomous cyclic engine) introduced in Sec.~\ref{sec:model} would naturally involve quantum effects in reality when we take atomic-scale systems. A direction for future works would be taking into account the genuine quantum effects~\cite{Scovil1959,Uzdin2015,KUP}. It would be also interesting to study the equivalence of the autonomous and non-autonomous models at the quantum level~\cite{KUP}.

\begin{acknowledgments}
We thank Hyun-Myung Chun, Jae Dong Noh, Hee Joon Jeon, and Sang Wook Kim for fruitful discussions and comments.
This research was supported by the NRF Grant No.~NRF-2017R1D1A1B03030872 (JU) and
2017R1D1A1B06035497 (HP), and 2018R1C1B5083863 (SHL).
\end{acknowledgments}

\appendix
\section{Global optimization}
\label{sec:series_expansion_procedure}

The global optimization condition, Eq.~\eqref{eq:first_derivative_condition}, leads to
\begin{subequations}
\begin{equation}
\displaystyle 1 - \frac{T_2 \ln\left[ (1-\epsilon^*)/\epsilon^* \right]}{T_1 \ln\left[ (1 - q^*)/q^* \right]} = \frac{q^* - \epsilon^*}{q^* (1-q^*) \ln\left[(1-q^*)/q^*\right]} \,,
\label{eq:optimal_condition_for_q}
\end{equation}
and
\begin{equation}
\displaystyle 1 - \frac{T_2 \ln\left[ (1-\epsilon^*)/\epsilon^* \right]}{T_1 \ln\left[ (1 - q^*)/q^* \right]} = \frac{(T_2 / T_1)(q^* - \epsilon^*)}{\epsilon^* (1-\epsilon^*) \ln\left[(1-q^*)/q^*\right]} \,.
\label{eq:optimal_condition_for_epsilon}
\end{equation}
\end{subequations}
By eliminating the left-hand side of Eqs.~\eqref{eq:optimal_condition_for_q} and \eqref{eq:optimal_condition_for_epsilon}, we obtain
the following simple relation
\begin{subequations}
\begin{equation}
\frac{T_2 q^* (1-q^*)}{T_1 \epsilon^* (1-\epsilon^*)} = 1 \,,
\label{eq:global_optimum_condition}
\end{equation}
or
\begin{equation}
\epsilon^* = \frac{1}{2} \left[ 1 -  U(\eta_C,q^*) \right] \,,
\label{eq:global_optimum_condition_solution}
\end{equation}
\end{subequations}
with
\begin{equation}
U(\eta_C,q^*) \equiv \sqrt{4\eta_C q^* (1-q^*) + (1 - 2q^*)^2} \,.
\label{eq:U_definition}
\end{equation}
By substituting $\epsilon^*$ as a function of $q^*$ in Eq.~\eqref{eq:global_optimum_condition_solution} to Eq.~\eqref{eq:optimal_condition_for_q} or Eq.~\eqref{eq:optimal_condition_for_epsilon},
we get the optimum condition
\begin{equation}
  \ln \left( \frac{\displaystyle 1-q^*}{\displaystyle q^*} \right) - \frac{\displaystyle T_2}{\displaystyle T_1} \ln \left[ \frac{\displaystyle 1 + U(\eta_C,q^*)}{\displaystyle 1 - U(\eta_C,q^*)} \right] =
  \frac{\displaystyle q^* - \frac{1}{2} + \frac{1}{2}U(\eta_C,q^*)}{\displaystyle q^*(1-q^*)} \,.
\label{eq:f_q_star}
\end{equation}
Furthermore, the condition in Eq.~\eqref{eq:f_q_star} leads to the following form of $\eta_\mathrm{op}$ from Eq.~\eqref{eq:eta},
\begin{equation}
\eta_\mathrm{op} = \frac{\displaystyle q^* - \frac{1}{2} + \frac{1}{2} U(\eta_C,q^*)}{\displaystyle q^* (1-q^*) \ln [(1-q^*)/q^*]} \,.
\label{eq:optimal_eta_for_q_star_T}
\end{equation}

In order to calculate the efficiency at maximum power for given $T_2/T_1$, first find the $q^*$ value satisfying Eq.~\eqref{eq:f_q_star} and substitute the $q^*$ value to Eq.~\eqref{eq:optimal_eta_for_q_star_T}.
As Eq.~\eqref{eq:f_q_star} is a transcendental equation, the closed-form solution for $\eta_\mathrm{op}$ is not possible in general.

We study analytically asymptotic behaviors of $\eta_\text{op}$ near $\eta_C=0$ and $\eta_C=1$.
First, examine the case for small $\eta_C$, using the series expansion of $q^*$ with respect to $\eta_C$ as
\begin{equation}
q^* = a_0 + a_1 \eta_C + a_2 \eta_C^2 + a_3 \eta_C^3 + \mathcal{O}\left({\eta_C^4}\right) \,.
\label{eq:q_series_expansion_wrt_eta_C}
\end{equation}
Substituting Eq.~\eqref{eq:q_series_expansion_wrt_eta_C} into Eq.~\eqref{eq:f_q_star} and expanding the equation with respect to  $\eta_C$ again, we obtain
\begin{equation}
c_1 \eta_C + c_2 \eta_C^2 + c_3 \eta_C^3 + \mathcal{O}\left({\eta_C^4}\right) = 0 \,,
\label{eq:f_star_expansition_wrt_eta_C}
\end{equation}
where $c_n$ is a function of a set of coefficients $\{a_0, \cdots, a_{n-1}\}$.
To satisfy Eq.~\eqref{eq:f_star_expansition_wrt_eta_C},
each $c_n$ should be identically zero. From $c_1=0$, we can easily find
\begin{equation}
\frac{2}{1-2a_0} = \ln \left( \frac{1-a_0}{a_0} \right) \,,
\label{eq:a_0_expression}
\end{equation}
from which we get $a_0\approx 0.083\ 222$. This serves as the lower bound of $q^*$.
From $c_2=0$ and $c_3=0$, we can express $a_1$ and $a_2$ in terms of $a_0$. From Eq.~\eqref{eq:global_optimum_condition_solution},
we can also find $\epsilon^*$ as $\epsilon^*=q^*-[a_0(1-a_0)/(1-2a_0)] \eta_C+\cdots$.

With the relations of coefficients in hand, we find the asymptotic behavior of $\eta_\mathrm{op}$ in Eq.~\eqref{eq:optimal_eta_for_q_star_T} by expanding it with respect to $\eta_C$ after substituting $q^*$ as the series expansion of $\eta_C$ in Eq.~\eqref{eq:q_series_expansion_wrt_eta_C}.
Then, we obtain the expression in Eq.~\eqref{eq:eta_op_expansion_wrt_eta_C_as_CA_main_text} in the main text,
\begin{equation}
\eta_\mathrm{op} = \frac{1}{2} \eta_C + \frac{1}{8} \eta_C^2 + \frac{7-24a_0+24a_0^2}{96(1-2a_0)^2} \eta_C^3 + \mathcal{O}\left({\eta_C^4}\right) \,.
\label{eq:eta_op_expansion_wrt_eta_C_as_CA}
\end{equation}
With this method, we are able to find the coefficients in terms of $a_0$ up to an arbitrary order in principle.

For $\eta_C \simeq 1$, we need to take into account a logarithmic singularity,
arising from $\ln [1-U(\eta_C,q^*)] \sim \ln (1-\eta_C)$ in Eq.~\eqref{eq:f_q_star}.
We take a singular series expansion of $q^*$ with respect to $1-\eta_C$ as
\begin{equation}
q^* = b_0 + b_1^\prime (1-\eta_C) \ln(1-\eta_C)+ b_1 (1-\eta_C) + \mathcal{O}\left[{(1-\eta_C)^2}\right] \,.
\label{eq:q_series_expansion_wrt_eta_C_near_one}
\end{equation}
Substituting Eq.~\eqref{eq:q_series_expansion_wrt_eta_C_near_one} into Eq.~\eqref{eq:f_q_star} and expanding the equation with respect to
$1-\eta_C$, we can identify the equation for $b_0$ as
\begin{equation}
\frac{1}{1-b_0} = \ln\left( \frac{1-b_0}{b_0} \right) \,,
\label{eq:qmax_solution}
\end{equation}
from which we get $b_0\approx 0.217\ 812$. This serves as the upper bound of $q^*$.
We also find $b_1^\prime = b_0 (1-b_0)^2$ and $b_1=b_0 (1-b_0)^2 ( 1 + \ln[b_0(1-b_0)])$.
Putting all these together into Eq.~\eqref{eq:optimal_eta_for_q_star_T}, we obtain
\begin{equation}
\begin{aligned}
\eta_\mathrm{op} = & 1 + (1-b_0) (1 - \eta_C) \ln (1 - \eta_C) \\
 & + (1-b_0) \ln[b_0(1-b_0)] (1 - \eta_C)
  + \mathcal{O}\left[ (1-\eta_C )^2\right] \,.
\end{aligned}
\label{eq:eta_op_expansion_near_eta_C_1}
\end{equation}


\begin{thebibliography}{00}
\bibitem{exp1} A similar result was reported experimentally in the lower right panel of Fig.~3(c) in~\cite{exp}, where, with $\eta_C\simeq 0.36$,
$\eta_\text{op}$ is about $25\%$ larger than $\eta_\text{CA}$ and the power output is about $65\%$ of its global maximum.
\end{thebibliography}
\end{document}